\RequirePackage{fix-cm}
\documentclass{svjour3}                     
\smartqed  
\usepackage{graphicx}
\newcommand{\be}{\begin{equation}}
\newcommand{\ee}{\end{equation}}
\def\tr{\mathop{\rm tr}\nolimits}

\def\ci{\mathop{\textrm{i}}\nolimits}

\begin{document}

\title{Type D vacuum solutions: a new intrinsic approach
}


\author{Joan Josep Ferrando        \and
   Juan Antonio S\'aez 
}


\institute{Joan Josep Ferrando \at
              Departament d'Astronomia i Astrof\'{\i}sica, Universitat
de Val\`encia, E-46100 Burjassot, Val\`encia, Spain. \\
                           \email{joan.ferrando@uv.es}           
           \and
          Juan Antonio S\'aez  \at
               Departament de Matem\`atiques per a l'Economia i l'Empresa,
Universitat de Val\`encia, E-46071 Val\`encia, Spain
}

\date{}

\maketitle

\begin{abstract}
We present a new approach to the intrinsic properties of the type D
vacuum solutions based on the invariant symmetries that these
spacetimes admit. By using tensorial formalism and without
explicitly integrating the field equations, we offer a new proof that the
upper bound of covariant derivatives of the Riemann tensor required
for a Cartan-Karlhede classification is two. Moreover we show that,
except for the Ehlers-Kundt's C-metrics, the Riemann derivatives
depend on the first order ones, and for the C-metrics they depend on
the first order derivatives and on a second order constant
invariant. In our analysis the existence of an invariant complex
Killing vector plays a central role. It also allows us to easily
obtain and to geometrically interpret several known relations. We
apply to the vacuum case the intrinsic classification of the type D
spacetimes based on the first order differential properties of the
2+2 Weyl principal structure, and we show that only six classes are
compatible. We define several natural and suitable subclasses and
present an operational algorithm to detect them.
\end{abstract}

\section{Introduction}

Type D vacuum solutions describe physically relevant gravitational
fields and an exhaustive geometrical comprehension of them is of
evident interest. The subfamily of static ones was obtained by
Ehlers and Kundt \cite{ehku} and the full set of type D vacuum
metrics by Kinnersley \cite{kin} using NP formalism. Recently, Edgar
et al. \cite{edgar-alfonso} have revisited the type D vacuum
metrics. They show that the GHP formalism is more suitable to obtain
several identities, say ${\cal I}$, which need a hard computer
support when using NP formalism \cite{Cz-Mc-a}, \cite{Cz-Mc-b}. They
also improve previous results \cite{aman}, \cite{colli1},
\cite{colli2} and show, without explicitly integrating the field
equations, that the Karlhede upper bound of covariant derivatives of
the Riemann tensor is two.

Here we analyze this subject by using a new approach that has the
following qualities: (i) We use plain tensorial formalism and all
our calculations are made without computer support. (ii) An
invariant complex Killing vector $Z$ plays a central role in our
study; in obtaining the Karlhede upper bound we show that $Z$ and
$\nabla Z$ determine, respectively, the first and the second
derivatives of the Riemann tensor, and for three different cases we
give the explicit expression of $\nabla Z$ in terms of $Z$ (in one
of the three cases $\nabla Z$ also depends on a constant second
order scalar invariant). (iii) We easily obtain identities ${\cal
I}$ and give an interesting geometric interpretation of them.

In this paper we also offer a generic and invariant classification
of the type  D vacuum metrics. Previously  known classifications
derive from the integration process of the vacuum equations
\cite{kin}, or from the study the Karlhede upper bound
\cite{edgar-alfonso}. Our invariant classification is generic
because it applies to the full set of type  D metrics, and it
imposes first order (tensorial) invariant conditions on the Weyl
tensor. We show here that the $2^{10}$ possible classes radically
decline to only 6 classes in the vacuum case.

In section \ref{sec-type-D} we present the notation used in the
paper and we write the Bianchi identities and the post-Bianchi
identities for a type D vacuum space-time. Moreover we introduce the
invariant complex Killing vector $Z$.

In section \ref{sec-identities} we point out the tensorial
invariants which collect the 0th- and 1st-order covariant
derivatives of the Riemann tensor: the 0th-order are given by the
algebraic Weyl invariants, namely, the complex eigenvalue $w$ and
the canonical bivector ${\cal U}$, and the 1st-order are given by
the complex Killing vector $Z$. By imposing the invariance of these
Riemann invariants along the invariant Killing vector $Z$ we easily
obtain several relevant identities, which are used in this paper. One
of them is identity ${\cal I}$ studied in \cite{edgar-alfonso} which
in our formalism admits a nice geometric interpretation: the
projections on the two Weyl principal planes of the real and
imaginary parts of $Z$ are collinear vectors.

The upper bound on the order of the Riemann covariant derivatives is
studied in section \ref{sec-bound}. The 2nd-order Riemann
derivatives are collected in the Killing 2--form $\nabla Z$, which
can be written in terms of $w$, ${\cal U}$, $Z$ and a complex scalar
$m$ as a consequence of the identities obtained in the previous section.
In the analysis of the scalar $m$ we distinguish three cases: the
Ehlers and Kundt C-metrics, the regular generalized C-metrics, and
the Kerr-NUT solutions (those type  D vacuum metrics that admit a
Killing tensor). In the last two cases we give the explicit
expression of $m$ in terms of $w$, ${\cal U}$, $Z$, and in the first
one we show that $m$ depends on these invariants and on a second
order constant invariant scalar.

In section \ref{sec-classification} we present the invariant
classification. Firstly we point out that two classifications of
type  D metrics based on first order (tensorial) invariant
conditions on the Weyl tensor can be considered \cite{fsD}. One of
them is the natural first order geometric classification of the 2+2
almost-product structure \cite{cf}, \cite{fs2+2} associated to the
Weyl tensor. The second one is defined by the first derivatives of
the Weyl eigenvalue $w$. We show that the $2^6$ classes of the first
classification notably decline to the 16 classes of the second one
when the Cotton tensor vanishes. Moreover, only 6 of these 16
classes are compatible with the vacuum condition. We finish this
section giving an operational algorithm to detect every class and
certain relevant subclasses, and pointing out the relationship with
previous classifications.

Section \ref{discussion} is devoted to discussing and commenting our
results. In appendix A  and appendix B we
prove some lemmas used in sections \ref{sec-bound} and
\ref{sec-classification}, respectively.

In this paper we work on an oriented space-time with a metric tensor
$g$ of signature $\{-,+,+,+\}$. The Riemann, Ricci and Weyl tensors
are defined as given in \cite{kramer} and are denoted, respectively,
by $Riem$, $Ric$ and $W$. For the metric product of two vectors we
write $(x,y) = g(x,y)$, and we put $x^2 = g(x,x)$. If $A$ and $B$
are 2-tensors, $A \cdot B$ denotes the 2-tensor $(A \cdot
B)^{\alpha}_{\ \beta} = A^{\alpha}_{\ \mu} B^{\mu}_{\ \beta}$, $A^2
= A \cdot A$, $A(x,y) = A_{\alpha \beta} x^{\alpha} y^{\beta}$, and
$(A,B) = \frac12 A_{\alpha \beta} B^{\alpha \beta}$.

\section{Type D vacuum metrics. Basic relations}
\label{sec-type-D}

A self--dual 2--form is a complex 2--form ${\cal F}$ such that
$*{\cal F}= \textrm{i}{\cal F}$, where $*$ denotes the Hodge dual
operator. We can associate biunivocally with every real 2--form $F$
the self-dual {2--form ${\cal
F}=\frac{1}{\sqrt{2}}(F-\textrm{i}*F)$. Here we refer to a
self--dual 2--form as a {\it self-dual bivector}. The endowed metric
on the 3-dimensional complex space of the self-dual bivectors is
${\cal G}=\frac{1}{2}(G-\textrm{i} \; \eta)$, $\eta$ being the
metric volume element of the space-time and $G$ the metric on the
space of 2--forms, $G_{\alpha \beta \gamma \delta} = g_{\alpha
\gamma} g_{\beta \delta} - g_{\alpha \delta} g_{\beta \gamma}$. The basic elements of self-dual bivector formalism and its relationship with the formalism based on orthonormal or on null tetrads can be found in  \cite{fms}.

Every double 2--form, and in particular the Weyl tensor $W$, can be
considered as an endomorphism on the space of the 2--forms. The
restriction of the Weyl tensor on the self-dual bivectors space is
the {\em self-dual Weyl tensor} and it is given by $ {\cal W} =
\frac12(W - \ci *W)$. The Petrov-Bel classification follows taking into account both the eigenvalue multiplicity and the degree of the minimal polynomial of this endomorphism. In \cite{fms} we have presented a complete study of this subject as well as the covariant determination of the geometric elements that appear in every class. 

In the case of a type  D spacetime the the self--dual Weyl tensor admits the
canonical expression:
\begin{equation}  \label{type-D-canonica}
{\cal W} = 2 w \, {\cal U} \otimes {\cal U} + w \, {\cal G}_{\perp}
= 3 w \, {\cal U} \otimes {\cal U} + w \, {\cal G}   \, ,
\end{equation}
where ${\cal U}$ is the {\em canonical bivector}, normalized
eigenbivector associated with the simple eigenvalue $-2w$, and
${\cal G}_{\perp} =  {\cal U} \otimes {\cal U} +  {\cal G}$ is its
orthogonal projector. If $\ell$ and $k$ are the Debever principal
null directions and $U = \ell \wedge k$, then ${\cal U} =
\frac{1}{\sqrt{2}}(U-\textrm{i}*U)$. The canonical bivector ${\cal
U}$  determines the two {\em principal planes} of a type D Weyl
tensor. The projector on the time-like (resp., space-like) principal
plane is $v = U^2$ (resp., $h = g-v = -(*U)^2 $), and $\Pi = v-h = 2
\, {\cal U} \cdot \bar{{\cal U}}$ is the {\em structure tensor}.
From now on $\bar{t}$ denotes the complex conjugate of a complex
tensor $t$.

The structure tensor $\Pi$ will play an important role here on. On one hand in section \ref{sec-bound} when studying the Cartan-Karlhede upper bound of type D vacuum solutions. On the other hand, in section \ref{sec-classification} when classifying this family of solutions. Indeed, the invariant decomposition of the
covariant derivative of the structure tensor $\Pi$ gives rise to the
classification of almost product  structures assumed in differential geometry \cite{naveira} \cite{gil-medrano}.  In \cite{fsD}, \cite{cf} and \cite{fs2+2} we have implemented these ideas to the general relativity framework and we have used them to classify type D metrics \cite{fsD}. In the case of 1+3 almost-product structures defined by a timeline unit vector this approach leads to the known concepts of acceleration, expansion, shear and vorticity. In the 2+2 considered case the geometric properties also have a kinematic interpretation \cite{cf}.

\subsection*{Bianchi identities}
\label{subsec-bianchi}
Under the vacuum condition $Ric =0$, Bianchi identities become
$\delta {\cal W} = 0$, where $(\delta {\cal W})_{\alpha \beta
\gamma} = - \nabla_{\lambda}{\cal W}^{\lambda}_{\ \alpha \beta
\gamma}$. From now on, we will write $\delta t = - \nabla \cdot t
\,$ for an arbitrary tensor $t$. Then, if we consider the canonical
expression (\ref{type-D-canonica}) of a type D Weyl tensor and we
take into account that $2 \, {\cal U}^2 = g$, Bianchi identities
write for the algebraic Weyl variables $\{w, {\cal U}\}$ as:
\be \label{bianchi}
\nabla {\cal U} = i(\delta {\cal U}){\cal G}_{\perp} \, , \qquad 3
\, i(\delta {\cal U}){\cal U} = \mbox{d} \ln w \, ,
\ee
where for a vector $X$ and a (p+1)-tensor $t$, $i(X)t$ denotes the
inner product, $[i(X)t]_{\underline{p}} = X^{\alpha}t_{\alpha
\underline{p}}$, the underline denoting multi-index.
\subsection*{Post-Bianchi identities}
\label{subsec-post-bianchi}
The integrability conditions of the second equation in
(\ref{bianchi}) lead to the post-Bianchi equations:
\be \label{post-bianchi}
\mbox{d} \chi  = 0 \, , \qquad \chi \equiv \, i(\delta {\cal U}){\cal U}
\, .
\ee

On the other hand, we can study the integrability conditions of the
first equation (\ref{bianchi}) by using the Ricci identities for the
canonical bivector ${\cal U}$,
\begin{equation} \label{ricci-U}
\nabla_{\alpha}  \nabla_{\beta} \,  {\cal U}_{\mu \nu} -
\nabla_{\beta} \nabla_{\alpha} \,  {\cal U}_{\mu \nu} = {\cal
U}_{\mu}^{\ \lambda} \, R_{\lambda \nu \beta \alpha} - {\cal
U}_{\nu}^{\ \lambda} \, R_{\lambda \mu \beta \alpha} \, .
\end{equation}
From the first condition in (\ref{bianchi}) and the canonical
expression (\ref{type-D-canonica}), these Ricci identities write:
\begin{eqnarray}
w = - ({\cal U}, T)\, , \qquad \quad \  T \cdot {\cal U} - {\cal U} \cdot T
= 0 \, ,
\label{ricci-U-1}\\
T \equiv \nabla \xi - \chi \otimes \xi \, , \qquad \xi \equiv \delta
{\cal U} \, , \qquad \chi \equiv \, i(\delta {\cal U}){\cal U} \, .
\label{T}  \label{ricci-U-2}
\end{eqnarray}
\subsection*{The invariant complex Killing vector}
\label{subsec-Killing}

Let $S$ be the symmetric part of $T$. The second equation in
(\ref{ricci-U-1}) implies $S \cdot {\cal U} - {\cal U} \cdot S = 0$.
On the other hand, if we develop equation (\ref{post-bianchi})
taking into account (\ref{bianchi}) we obtain $S \cdot {\cal U} +
{\cal U} \cdot S = 0$, and then $S = 0$:
\begin{equation} \label{pre-Killing}
{\cal L}_{\xi} g - \xi \tilde{\otimes} \chi = 0 \, ,
\end{equation}
where ${\cal L}_{\xi}$ denotes the Lie derivative with respect the
field $\xi$. From this condition and the second equation in
(\ref{bianchi}), we recover the following result \cite{hs2},
\cite{fsEM-sym}:
\begin{proposition} \label{prop-Killing}
The type D vacuum solutions admit the invariant complex Killing
vector
\begin{equation} \label{Killing}
Z = w^{-\frac13} \xi \, , \qquad  \xi \equiv \delta {\cal U} \, .
\end{equation}
\end{proposition}

There is another useful consequence of the Bianchi and post-Bianchi
identities. Indeed, being $Z$ a Killing vector, (\ref{ricci-U-2})
and (\ref{Killing}) imply $\nabla  Z = \mbox{d} Z = w^{-\frac13} T$, and
from (\ref{ricci-U-1}) we obtain that the Killing 2-form $\nabla
Z$ satisfies:
\begin{equation} \label{Killing-2-form-a}
({\cal U}, \nabla Z) = - w^{\frac23} \, , \qquad   \nabla Z \cdot
{\cal U}- {\cal U} \cdot \nabla Z = 0 \, .
\end{equation}

The invariant Killing vector $Z$ also exists in the charged
counterpart of the type  D vacuum metrics. For an in-depth study of this
and other invariant symmetries and their close relation with the
curvature tensor see \cite{fsEM-sym} and references therein.

\section{Some relevant identities: obtaining and interpretation}
\label{sec-identities}

The Bianchi equations (\ref{bianchi}) and definition (\ref{Killing})
imply that the first derivatives of the algebraic Weyl variables
$\{w, {\cal U}\}$ depend on the Killing vector $Z$. Then, we obtain
the following set of zero, first and second order independent
Riemann derivatives:
\begin{enumerate}
\item[] {\it $0$th-order}: $w, {\cal U}$.
\item[] {\it $1$st-order}: $Z = w^{-\frac13} \delta {\cal U}$
\item[] {\it $2$nd-order:} $\nabla Z$.
\end{enumerate}
Moreover, in terms of  $w$, ${\cal U}$ and $Z$ the covariant
derivatives of  $\{w, {\cal U}\}$ write:

\be \label{dw-dU}
\mbox{d} w = 3 w^{\frac43} i(Z){\cal U} \, , \qquad \nabla {\cal U} =
w^{\frac13}i(Z)({\cal U} \otimes {\cal U} + {\cal G}) \, .
\ee
On the other hand, the integrability condition for the Killing
equation holds: $\nabla \nabla Z = i(Z)Riem$. Then, we have for the
{\it $3$rd-order} Riemann derivatives:

\begin{equation} \label{ddZ}
\nabla \nabla Z = w^{-\frac23}i(Z)(3 {\cal U} \otimes
{\cal U} + {\cal G}) .
\end{equation}
Consequently, we obtain the following result:
\begin{proposition} \label{proposition-2}
All the covariant derivatives of the Riemann tensor of a type D
vacuum solution depend at most on the second order ones.
\end{proposition}

Now we study new restrictions on the Killing 2-form which allow us
to improve the above result. The invariant vector $Z$ being a
complex Killing vector, we have ${\cal L}_Z Riem = 0$, ${\cal L}_Z
\nabla Riem = 0$ and, consequently, similar relations hold for the
Riemann invariants $\{w, {\cal U}, Z\}$ and their complex conjugate
invariants:\\[3mm]
$\circ \ {\cal L}_Z w = 0$.
That is $(Z, \mbox{d} w) =0$. This condition is an identity as a
consequence of the first expression in (\ref{dw-dU}).\\[3mm]
$\circ \ {\cal L}_{Z} \bar{w} = 0$.
That is $({\bar{Z}, \mbox{d} w) =0}$. From the first expression in
(\ref{dw-dU}) and definition (\ref{Killing}), this condition can be
written as one of the following equivalent conditions:
\be \label{x-identities-a}
{\cal U}(Z,\bar{Z}) = 0 \, , \qquad \quad \ \ {\cal U}(\delta{\cal
U}, \delta \bar{{\cal U}})=0 \, .
\ee
The real and imaginary parts of the second condition in
(\ref{x-identities-a}) give the scalar identities:
\be \label{x-identities-b}
U(\delta U, \delta \! *\!U)=0 \, , \qquad *U(\delta U, \delta
\!*\!U)=0 \, .
\ee
If we write these identities in NP formalism we obtain:
\be \label{x-identities-c}
\pi \bar{\pi} - \tau \bar{\tau} = 0  \, , \qquad \quad \ \ \rho
\bar{\mu} - \bar{\rho} \mu = 0  \, .
\ee
These restrictions on the NP coefficients were already known, and
the difficulties in obtaining them have been outlined. Czapor and
McLenaghan \cite{Cz-Mc-a}, \cite{Cz-Mc-b}  showed
(\ref{x-identities-c}) using computer support and they claimed that
their calculations required pages and ``would be virtually
impossible by hand". Recently \cite{edgar-alfonso} their obtaining
has been widely improved by using the GHP formalism.

Here, we have used plain tensorial formalism to obtain
(\ref{x-identities-b}). Our approach has several qualities. Firstly,
the calculation is simple and straightforward and highlights the
meaning of the identities: they state that the Weyl eigenvalue $w$
is invariant under the invariant complex Killing vector $\bar{Z}$.
Secondly, it offers a nice geometric interpretation. Indeed,
(\ref{x-identities-b}) equivalently writes as:
\be \label{x-identities-d}
v(\delta U) \wedge v(\delta \! *\!U)=0 \, , \qquad h(\delta U)
\wedge h(\delta \!*\!U)=0 \, ,
\ee
conditions which state that the projections of the first order
invariant vectors $\delta U$ and $\delta \! * \! U$ on the two Weyl
principal planes are collinear. Thirdly, a similar geometric
interpretation can be obtained in terms of the invariant Killing
vector. In fact, the equivalent condition given by the first
expression in (\ref{x-identities-a}) writes $U(Z_1,Z_2)=
*U(Z_1,Z_2)=0$, where $Z= Z_1 + \ci Z_2$.  Then
(\ref{x-identities-b}) is equivalent to:
\be \label{x-identities-e}
v(Z_1) \wedge v(Z_2)=0 \, , \qquad h(Z_1) \wedge h(Z_2)=0 \, ,
\ee
conditions which state that the projections of the invariant Killing
vectors $Z_1$ and $Z_2$ on the two Weyl principal planes are
collinear.\\[3mm]
$\circ \ {\cal L}_Z {\cal U} = 0$.
As a consequence of the second expression in (\ref{dw-dU}), this
condition is equivalent to the second constraint for the Killing
2-form given in (\ref{Killing-2-form-a}).\\[3mm]
$\circ \ {\cal L}_{Z} \bar{{\cal U}} = 0$.
Taking into account the second expression in (\ref{dw-dU}), this
condition can be written as the following constraint for the Killing
2-form:
\begin{equation} \label{Killing-2-form-b}
\nabla Z \cdot \bar{{\cal U}}- \bar{{\cal U}} \cdot \nabla Z =
\bar{w}^{\frac13}\, \bar{{\cal G}}(Z \wedge \bar{Z}) \, ,
\end{equation}
where, for a double 2-form $V$ and a 2-form $F$, $V(F)$ denotes the
action of $V$ on $F$ as an endomorphism, $V(F)_{\alpha \beta} =
\frac12 V_{\alpha \beta}^{\ \ \gamma \delta} F_{\gamma \delta}$.
\ \\
$\circ \ {\cal L}_{Z} \bar{Z} = 0$.
This condition restricts the electric part of the Killing
2-form with respect the invariant Killing vector:
\be \label{Lie-Z}
i(Z) \nabla \bar{Z} - i(\bar{Z}) \nabla Z = 0 \, .
\ee
%

\section{Upper bound on the order of the Riemann covariant derivatives}
\label{sec-bound}
We have seen in the previous section that the second derivatives of
the Riemann tensor depend on the Killing 2-form $\nabla Z$, which
are restricted by conditions (\ref{Killing-2-form-a}),
(\ref{Killing-2-form-b}) and (\ref{Lie-Z}}). The first condition in
(\ref{Killing-2-form-a}) gives the ${\cal U}$-component of $\nabla
Z$, and the second condition in (\ref{Killing-2-form-a}) implies
that its self-dual part orthogonal to ${\cal U}$ vanishes, ${\cal
G}_{\perp} (\nabla Z) = 0$. On the other hand, condition
(\ref{Killing-2-form-b}) determines the anti-self-dual part of
$\nabla Z$ which is orthogonal to $\bar{{\cal U}}$, $\bar{{\cal
G}}_{\perp} (\nabla Z) = \bar{w}^{\frac13} \bar{{\cal G}}(Z \wedge
\bar{Z}) \cdot \bar{{\cal U}}$. Consequently, we obtain the
following expression for the Killing 2-form $\nabla Z$:
\begin{equation} \label{killing-2-form-c}
\nabla Z = w^{\frac23}\,{\cal U} + m  \, \bar{{\cal U}} +
\bar{w}^{\frac13} \, \bar{{\cal G}}(Z \wedge \bar{Z}) \cdot
\bar{{\cal U}} \, .
\end{equation}

Note that expression (\ref{killing-2-form-c}) gives $\nabla Z$ in
terms of 0th- and 1st-order Riemann derivatives ($w$, ${\cal U}$ and
$Z$) and a complex scalar $m$. Moreover, if we put expression
(\ref{killing-2-form-c}) in (\ref{Lie-Z}) we obtain:
\vspace{-3mm}
\begin{equation} \label{m-eq}
 \mu \Pi(\bar{Z}) - \bar{\mu} Z = \bar{\nu}  \Pi(Z) -
\nu \bar{Z} \, , \ \ \ \mu \equiv m + \frac12 \bar{w}^{\frac{1}{3}}
(Z, \bar{Z}) , \ \ \
 \nu \equiv w^{\frac23} + \frac12 w^{\frac13} (Z,Z) \, .
  \end{equation}

Thus, we have a system of linear equations for the 2nd-order
differential scalar $m = -(\bar{\cal U}, \nabla Z)$, with
coefficients depending on 0th- and 1st-order Riemann derivatives. It
is worth remarking that the first derivatives of $m$ depend on
1st-order Riemann derivatives. Indeed, from (\ref{dw-dU}),
(\ref{ddZ}) and (\ref{killing-2-form-c}) we obtain:
\begin{equation} \label{dm}
vspace{-2mm}
2 \, \mbox{d} m = [4 \bar{w} + \bar{w}^{2/3}(\bar{Z},\bar{Z})]\, i(Z)
\bar{{\cal U}} - \bar{w}^{2/3} (Z, \bar{Z})\, i(\bar{Z}) \bar{{\cal
U}} \, .
\end{equation}

Now, we study whether system (\ref{m-eq}) allows us to obtain $m$ in
terms of $w$, ${\cal U}$ and $Z$. We consider three cases and we
will make use of the following result which is proven in appendix
A:
\begin{lemma} \label{lemma-ZZ-nonnull}
For any type  D vacuum solution the invariant Killing vector $Z$ is
a non null vector, $(Z,Z) \not=0$.
\end{lemma}

\vspace{-10mm}
\subsection{Upper bound for the Kerr-NUT solutions}
\label{subsec-boundKN}
vspace{-3mm}
Hougston and Sommers \cite{hs1} showed that, with the exception of
the generalized $C$-metrics, the other type  D vacuum solutions
admit a Killing tensor. {\it We call Kerr-NUT metrics the type  D vacuum
solutions where such a Killing tensor exists} \cite{fsEM-sym}. In a
subsequent paper Hougston and Sommers \cite{hs2} showed that the
complex Killing vector $Z$ degenerates (it defines a unique Killing
direction) if, and only if, the metric is a Kerr-NUT solution. This
result can be stated in the following terms:
\vspace*{-3mm}
\begin{lemma} \label{lemma-Kerr-NUT}
The Kerr-NUT vacuum metrics are the type  D vacuum  solutions  such that
the complex Killing vector $Z$ given in (\ref{Killing}) satisfies $Z
\wedge \bar{Z}=0$, or equivalently, $\bar{Z} = e^{\kappa   {\ci}} Z$,
where $\kappa$ is a real number.
\end{lemma}
\vspace*{-3mm}

This lemma and expression (\ref{killing-2-form-c}) imply that the
Killing 2-form $\nabla Z$ is aligned with the Weyl principal
structure, $\nabla Z = w^{\frac23}\,{\cal U} + m \, \bar{{\cal U}}
$, in accordance with a known result \cite{fsEM-sym}. From here and
taking into account that $\bar{Z} = e^{\kappa \ci} Z$ we have $m =
\bar{w}^{\frac23} e^{-\kappa \ci}$. Moreover lemma
\ref{lemma-ZZ-nonnull} implies in this case $(Z, \bar{Z}) \not=0$,
and we obtain the following expression for the $2$nd-order Riemann
derivatives:
\vspace*{-3mm}
\begin{equation} \label{nablaZ-Kerr-NUT}
 \displaystyle \nabla Z =
w^{\frac23}\, {\cal U} + \frac{(Z,Z)}{(Z,\bar{Z})}
\bar{w}^{\frac23}\, \bar{{\cal U}}
 \, .
\end{equation}
\vspace*{-1mm}
Consequently, we have the following result:
\begin{theorem} \label{theorem-1}
All the covariant derivatives of the Riemann tensor of a Kerr-NUT
vacuum solution depend at most on the first order ones.
\end{theorem}

\vspace{-3mm}

\subsection{Upper bound for the regular generalized $C$-metrics}
\label{subsec-bound-rC}

The type  D vacuum solutions which are not Kerr-NUT metrics have
been named {\it generalized $C$-metrics}. As a consequence of lemma
\ref{lemma-Kerr-NUT} they can be characterized by the condition $Z
\wedge \bar{Z}\not=0$. In these solutions the projection on the
time-like principal plane $v(Z)$ of the complex Killing vector $Z$
is necessarily a non null vector: $v( Z,Z) \not=0$. This fact is a
consequence of the following lemma which is proven in appendix
A:

\begin{lemma} \label{lemma-null}
The type  D vacuum solutions that satisfy $v(Z , Z)=0$ are
Kerr-NUT metrics, that is $Z \wedge \bar{Z}=0$.
\end{lemma}

Before studying linear system (\ref{m-eq}) for the case  $Z \wedge
\bar{Z}\not=0$ we must consider the following result which is proven
in appendix A:

\begin{lemma} \label{lemma-Cmetrics}
The strict (Ehlers and Kundt) $C$-metrics are the type  D vacuum
solutions that satisfy $Z \wedge \bar{Z}\not=0$ and $Z \wedge
\Pi(\bar{Z})=0$. Moreover, in this case $\Pi (\bar{Z}) = Z$,
$\bar{m} = m$, $\bar{w} = w$.
\end{lemma}

We consider the strict $C$-metrics in the following section. Now we
study the complementary set, the {\it regular generalized
$C$-metrics}, which can be characterized by the conditions $Z \wedge
\bar{Z}\not=0$ and $Z \wedge \Pi(\bar{Z})\not=0$ as a consequence of
lemmas \ref{lemma-Kerr-NUT} and \ref{lemma-Cmetrics}.  Under these
constraints the system (\ref{m-eq}) for the scalar $m$ admits a
solution given by:
\begin{eqnarray}
m =  m(w, {\cal U}, Z) \equiv -\frac12 (Z, \bar{Z}) \bar{w}^{\frac13}  +
\frac{D}{\Delta}\, , \quad  \Delta \equiv (Z, Z) (\bar{Z}, \bar{Z}) -
[\Pi(Z, \bar{Z})]^2 \, , \label{m-Cg-1} \qquad \\[2mm]
 D \equiv  \Pi(Z,\bar{Z}) [\nu\, (Z,\bar{Z}) -
\bar{\nu} \, \Pi(Z,Z)] - (Z,Z) [ \nu\, \Pi(\bar{Z},\bar{Z}) -
\bar{\nu} \, (Z,\bar{Z})]      \, . \qquad \, \label{m-Cg-2}
\end{eqnarray}
where $\nu$ is the scalar given in (\ref{m-eq}). Note that $\Delta
\not= 0$ for a regular generalized C-metric.  Indeed, the first
order Riemann scalar $\Delta$ is the square of the 2-form $Z
\wedge \Pi(\bar{Z}) \not=0$. Thus, $\Delta =0$ states that $Z \wedge
\Pi(\bar{Z})$ is a null 2-form, and the constraint
(\ref{x-identities-a}) implies $v(Z,Z)=0$, which is not possible as
a consequence of lemma \ref{lemma-null}.

Thus, the Killing 2-form $\nabla Z$ takes the expression
(\ref{killing-2-form-c}) where $m$ depends on first and second
Riemann derivatives as (\ref{m-Cg-1}-\ref{m-Cg-2}). Consequently, we
obtain the following result:
\begin{theorem} \label{theorem-2}
All the covariant derivatives of the Riemann tensor of a regular
generalized C-metric depend at most on the first order ones.
\end{theorem}

\subsection{Upper bound for the strict (Ehlers and Kundt) $C$-metrics}
\label{subsec-sC}

At this point, we still have to study the case $Z \wedge
\bar{Z}\not=0$, $Z \wedge \Pi(\bar{Z})=0$ which corresponds to the
strict C-metrics as a consequence of lemma \ref{lemma-Cmetrics}.
This lemma also states that $\Pi (\bar{Z}) = Z$, $\bar{m} = m$,
$\bar{w} = w$. Under these constraints (\ref{m-eq}) becomes an
identity and it does not allow us to obtain the scalar $m$ in terms of
0th- and 1st-order Riemann derivatives.

Let us consider the 2nd-order Riemann scalar:
\begin{equation} \label{K}
K \equiv [(Z,\bar{Z}) -2 m w^{-\frac13}]^2 + 12 [w^{\frac13} (Z,Z) +
w^{\frac23}] \, , \quad m \equiv -(\bar{\cal U}, \nabla Z) \, .
\end{equation}

From (\ref{dw-dU}), (\ref{killing-2-form-c}) and (\ref{dm}) we
obtain $\mbox{d} K = 0$. Therefore $m$, and thus $\nabla Z$, can be
obtained in terms of $w$, $Z$ and the invariant scalar $K$.
Consequently, we arrive at the following result:
\begin{theorem} \label{theorem-3}
All the covariant derivatives of the Riemann tensor of a strict
(Ehlers and Kundt) C-metric depend at most on the first order ones
and on a second order constant invariant.
\end{theorem}
\vspace{-5mm}
\section{Invariant classification}
\label{sec-classification}

In studying the Karlhede upper bound of derivatives of the Riemann
tensor in the above section, we have considered three classes of type
D vacuum solutions. Every class has specific geometric properties
and admits an invariant characterization in terms of 1st-order
Riemann derivatives, that is, in terms of the invariant
Killing vector $Z$:
\begin{enumerate}
\item[] {\it Kerr-NUT solutions}: $Z \wedge \bar{Z} = 0$.
\item[] {\it Strict (Ehlers and Kundt) $C$-metrics}:
$Z \wedge \bar{Z} \not= 0$, $\ Z \wedge \Pi(\bar{Z}) = 0$.
\item[] {\it Regular generalized $C$-metrics}: $Z \wedge
\bar{Z} \not= 0$, $\ Z \wedge \Pi(\bar{Z}) \not= 0$.
\end{enumerate}
Previous classifications have been introduced by Kinnersley
\cite{kin} in integrating the Einstein vacuum solutions or by Edgar
et al. \cite{edgar-alfonso} in studying the Karlhede upper bound.

Our aim here is to present a classification of generic type  D
spacetimes which is not induced by the Einstein field equations, and
subsequently to study the classes which are compatible with the vacuum
condition. This generic invariant classification can offer new geometrical and/or physical insight provided that it is defined by specific geometrical and/or physical restrictions.

\vspace{-4mm}

\subsection{Classifying  D metrics}
\label{subsec-class-D}

The Weyl tensor of a type  D metric determines a 2+2 almost product
structure defined by the principal 2-planes, and two real scalars
defined by the complex eigenvalue. Then, naturally, we can
consider two different classifications defined by first order
differential conditions.

The first one corresponds to the classification of the 2+2 principal
structure taking into account the invariant decomposition of the
covariant derivative of the structure tensor $\Pi$ or, equivalently,
according to the foliation, minimal or umbilical character of each
principal plane. This classification is assumed in differential geometry \cite{naveira} \cite{gil-medrano} and has been adapted to the general relativity framework \cite{fsD}, \cite{cf}, \cite{fs2+2}.  In \cite{fsD} we give the following.

\begin{definition}
Taking into account the foliation, minimal or umbilical character of
each principal $2$-plane we distinguish $2^6=64$ different classes
of type  D metrics.

We denote the classes as {\rm D}$^{p \, q \,r}_{lmn}$, where the
superscripts $p,q,r$ take the value $0$ if the time-like principal
plane is, respectively, a foliation, a minimal or an umbilical
plane, and they take the value $1$ otherwise. In the same way, the
subscripts $l, m , n$ collect the foliation, minimal or umbilical
nature of the space-like principal plane.
\end{definition}

The most degenerated class that we can consider is D$^{000}_{000}$
which corresponds to a product structure, and the most regular one
is D$^{111}_{111}$ which means that neither the time-like plane nor
the space-like plane are foliation, minimal or umbilical planes. We
will put a dot in place of a fixed script (1 or 0) to indicate the
set of metrics that cover both possibilities. So, for example, the
metrics of type D$^{111}_{11 \, \cdot}$ are the union of the classes
D$^{111}_{111}$ and D$^{111}_{110}$; or a metric is of type D$^{0 \,
\cdot \, \cdot}_{\, \cdot \, \cdot \, \cdot}$ if the time-like
2--plane is a foliation. Type D$^{\cdot \, \cdot \, 0}_{\cdot \,
\cdot \, 0}$ corresponds to an umbilical structure (both planes are
umbilical).

Here we use the following first order concomitants of the canonical
bivector ${\cal U} = \frac{1}{\sqrt{2}}(U-\textrm{i}*U)$ to
characterize some of these classes:
\begin{equation} \label{Sigma-Phi-Psi}
 \Sigma  \equiv  \nabla {\cal U} - i(\delta {\cal U}){\cal
G}_{\perp} , \quad \Phi \equiv i(\delta U) U - i(\delta \!*\!U)
\!*\!U , \quad \Psi \equiv -i(\delta U) \!*\!U - i(\delta \!*\!U) U
.
\end{equation}
More precisely, we have the following results \cite{fsD},
\cite{fs2+2}, \cite{fs-two}:
\begin{lemma}
\label{lemma-class-D}
In a type  D space-time,

\noindent (a) The principal time-like plane is a foliation (type
{\rm D}$^{0 \, \cdot \, \cdot }_{\, \cdot \, \cdot \, \cdot}$) if,
and only if, $h(\Psi)=0$.

\noindent(b) The principal space-like plane is a foliation (type
{\rm D}$^{\, \cdot \, \cdot \, \cdot}_{0 \,\cdot \, \cdot}$) if, and
only if, $v(\Psi)=0$.

\noindent (c) The principal time-like plane is minimal (type {\rm
D}$^{\cdot \, 0 \, \cdot}_{\cdot \;  \cdot \; \cdot}$) if, and only
if, $h(\Phi)=0$.

\noindent (d) The principal space-like plane is minimal (type {\rm
D}$^{\cdot \,\; \cdot \, \; \cdot}_{\cdot \; 0 \; \cdot}$) if, and
only if, $v(\Phi)=0$.

\noindent (e) The principal structure is umbilical (type {\rm
D}$^{\cdot \, \cdot \, 0}_{\cdot \, \cdot \, 0}$) if, and only if,
$\Sigma = 0$.

\end{lemma}
The physical meaning of the types in lemma above follows from the kinematic interpretation of the geometric conditions that define them \cite{fsD} \cite{cf}. The umbilical condition $\Sigma=0$ implies that the two Debever null directions (lying on the time-like principal plane) are shear-free geodesics. The minimal or foliation character of the space like plane state, respectively, that both Debever null directions are expansion-free or vorticity-free. Moreover, the vectors $h(\Phi)$ and $h(\Psi)$ have been named, respectively, the expansion and the rotation of the timeline plane \cite{cf}. Similarly, the vectors $v(\Phi)$ and $v(\Psi)$ are, respectively, the expansion and the rotation of the spacelike plane.

The second classification that we can consider is defined by first
derivatives of the Weyl eigenvalue $w$ \cite{fsD}:

\begin{definition}
Let $w = e^{\frac32( \phi + \ci \psi)}$ be the Weyl eigenvalue.
Taking into account the relative position between the gradients
$\mbox{\rm d} \phi$, $\mbox{\rm d} \psi$ and each principal
$2$-plane we distinguish $2^4=16$ different classes of type  D
metrics.

We denote the classes as {\rm D}$[p \, q , r s]$, where $p,q,r,s$
take the value $0$ if, respectively, the 1-form $v(\mbox{\rm d}
\psi)$, $v(\mbox{\rm d} \phi)$, $h(\mbox{\rm d} \psi)$, $h(\mbox{\rm
d} \phi)$ vanishes, and they take the value $1$ otherwise.
\end{definition}

The most degenerated class D$[00,00]$ is covered by the type  D
metrics with constant eigenvalues, and the most general one
D$[11,11]$ by those type  D spacetimes for which the gradients of
both, the modulus and the argument of the Weyl eigenvalue, have non
zero projection onto the principal planes. As above, a dot means
that a condition is not fixed. A constant modulus, $\mbox{d}
\phi=0$, corresponds to type D$[\, \cdot 0, \, \cdot 0]$, and a
constant argument, $\mbox{d} \psi =0$, corresponds to the metrics of
type D$[0 \, \cdot ,0\, \cdot ]$.

The two invariant classifications of type  D metrics presented
above have not, {\it a priori}, any relationship. Consequently, they
define $2^{10}$ different classes. Nevertheless, we will see that
the Einstein field equations or other restrictions on the Ricci
tensor forbid many of these classes and correlate both classifications.

\vspace*{-3mm}
\subsection{The sixteen classes of type D metrics with vanishing
Cotton tensor} \label{subsec-class-Cotton}

The spacetime Cotton tensor $P$ depends on the Ricci tensor as
$P_{\mu \nu , \beta} \equiv \nabla_{[\mu} Q_{\nu] \beta}$, $2Q
\equiv Ric - \frac{1}{6} (\tr Ric) g$. The Bianchi identities equal
the Cotton tensor with the divergence of the Weyl tensor.
Consequently, if the Cotton tensor of a type  D metric vanishes the
Bianchi identities take the same expression (\ref{bianchi}) as the
vacuum case.

From the point (v) in lemma \ref{lemma-class-D} the first condition
in (\ref{bianchi}) means that the principal structure is umbilical
(the principal directions are shear free null geodesics accordingly
to the Goldberg-Sachs theorem), that is, the space-time is of type
D$^{\cdot \cdot 0}_{\cdot \cdot 0}$. On the other hand, we have
$\Phi + \ci \Psi = 2 \chi$, and the second equation in
(\ref{bianchi}) is equivalent to:
\begin{equation} \label{rainich}
\Phi =  \mbox{d} \phi \ ; \qquad \qquad \Psi = \mbox{d} \psi \, .
\end{equation}
Note that in accordance with the Rainich theorem \cite{rai}, this
last equation states that the Weyl principal planes define a
Maxwellian structure \cite{fsD}, \cite{fs2+2}, \cite{fs-two}.

On the other hand, (\ref{rainich}) and lemma \ref{lemma-class-D}
imply that the modulus and the argument of the Weyl eigenvalue
govern, respectively, the minimal and the foliation character of the
principal planes. This relation establishes a bijection between the
classes of the two classifications that we have presented above.
More precisely, we have \cite{fsD}:

\begin{theorem} \label{theorem-cotton}
Every type  D spacetime with zero Cotton tensor is of type
{\rm D}$^{\cdot\; \cdot\; 0}_{\cdot \; \cdot \; 0}$. Moreover, it is of
class {\rm D}$^{ p\; q \;0}_{l \; m \;0}$ if, and only if, it is of
class {\rm D}$[lm,pq]$. So we have just 16 classes of type D spacetimes
with zero Cotton tensor.
\end{theorem}
\vspace{-6mm}

\subsection{The six classes of type D vacuum solutions}
\label{subsec-class-D-vacuum}

\vspace*{-3mm}
As a consequence of the above theorem there are at most sixteen
classes {\rm D}$[lm,pq]$ of type  D vacuum solutions. Now we show
that only six of these classes are compatible with the vacuum
equations. This result is based on the following lemma which is
proven in appendix B:

\begin{lemma}  \label{lemma-Phi-Psi}
In a type  D vacuum solution:

(i) If $v(\Phi) = 0$ then $v(\Psi)=0$.

(ii) If $h(\Phi) = 0$ then $h(\Psi)=0$.

(iii) If $v(\Psi) = 0$ then either $v(\Phi)=0$ or $h(\Psi) = 0$.

(iv) If $h(\Psi) = 0$ then either $h(\Phi)=0$ or $v(\Psi)=0$.
\end{lemma}

Point (i) of this lemma implies that if a metric is of type {\rm
D}$[\cdot 0, \cdot \cdot]$ then it is of type {\rm D}$[0 0, \cdot
\cdot]$. Consequently, type {\rm D}$[1 0, \cdot \cdot]$ is
forbidden. Similarly, point (ii) implies that type {\rm D}$[\cdot
\cdot, 1 0]$ is forbidden. On the other hand, point (iii) states
that type {\rm D}$[0 \cdot, \cdot \cdot]$ implies either type {\rm
D}$[0 0, \cdot \cdot]$ or type {\rm D}$[0 \cdot, 0 \cdot]$.
Consequently, class {\rm D}$[0 1, 1 1]$ is forbidden. Similarly
class {\rm D}$[1 1, 0 1]$ is forbidden as a consequence of point
(iv). Finally, a metric of class {\rm D}$[0 0, 0 0]$ is a product
metric which in the vacuum case implies a flat spacetime.

The remaining six compatible classes are presented below in a flow
chart which illustrate the degeneration paths from some classes to
others. Note that no classes in the first degeneration level are
compatible with the vacuum condition: the regular class can decline
to three of the six classes in the second level. In the third level
only two of the four classes are compatible.

Classes {\rm D}$[0 1, 0 0]$, {\rm D}$[0 0, 0 1]$ and {\rm D}$[0 1, 0
1]$ have Weyl real eigenvalues and correspond, respectively, to the
A-metrics, B-metrics and C-metrics by Ehlers and Kundt \cite{ehku}.
Classes {\rm D}$[1 1, 0 0]$ and {\rm D}$[0 0, 1 1]$ are the NUT-like
generalization of the A-metrics and B-metrics, respectively. The
regular class {\rm D}$[1 1, 1 1]$ contains both the regular
generalized C-metrics and the regular Kerr-NUT metrics. We can
summarize these results in the following.

 \setlength{\unitlength}{0.9cm} 
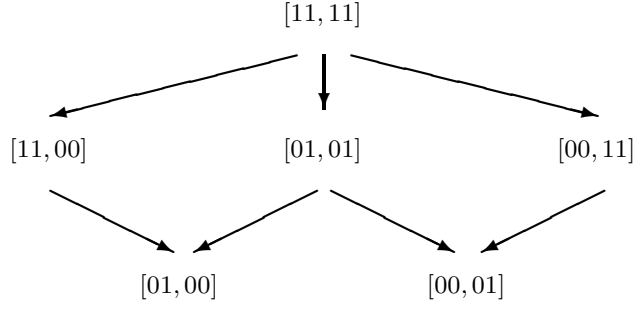
{
\noindent
\begin{picture}(5,5)(-2,13.5)
\thicklines

\put(6,17.5){\vector(4,-1){3.6 }}
\put(5.2,17.5){\vector(-4,-1){3.6}}

\put(5.6,17.5){\vector(0,-1){0.8 }}

\put(5,18){$ [11,11]$} \put(5,16){$ [01,01]$} \put(1,16){$ [11,00]$}
\put(9,16){$ [00,11]$}

\put(1.6,15.5){\vector(2,-1){1.8 }}
\put(5.5,15.5){\vector(-2,-1){1.8 }}

 \put(5.7,15.5){\vector(2,-1){1.8 }}
\put(9.7,15.5){\vector(-2,-1){1.8 }}

 \put(7.1,14){$ [00,01]$} \put(2.9,14){$ [01,00]$}

\end{picture}

}

\vspace{-3mm}

\begin{theorem} \label{theorem-vacuum}
Taking into account the first derivatives of the Weyl tensor we can
consider six classes of type  D vacuum solutions which can be
characterized by the following conditions:
\begin{enumerate}
\item[] {\rm D}$[0 1, 0 0]$ ({\it A-metrics}): $h(\Phi) = 0$, $v(\Psi)=0$.
\item[] {\rm D}$[0 0, 0 1]$ ({\it B-metrics}): $v(\Phi) = 0$, $h(\Psi)=0$.
\item[] {\rm D}$[0 1, 0 1]$ ({\it C-metrics}): $v(\Phi) \not= 0$, $h(\Phi)
\not= 0$, $v(\Psi)=0$ $h(\Psi)=0$.
\item[] {\rm D}$[1 1, 0 0]$ ({\it A-NUT-metrics}): $h(\Phi)= 0$, $v(\Psi)\not=0$.
\item[] {\rm D}$[0 0, 1 1]$ ({\it B-NUT-metrics}): $v(\Phi)= 0$, $h(\Psi)\not=0$.
\item[] {\rm D}$[1 1, 1 1]$ ({\it Regular C and Kerr-NUT metrics}):
$v(\Psi)\not= 0$, $h(\Psi)\not=0$.
\end{enumerate}
\end{theorem}
\vspace{-4mm}
\subsection{Some relevant subclasses and a summary in algorithmic form}

As showed in section \ref{sec-bound} condition $Z \wedge \bar{Z}  =
0$ characterizes the Kerr-NUT metrics. In terms  of the vectors
$\Phi$ and $\Psi$ this condition writes $N = 0$, where
\begin{equation} \label{N-KN}
N \equiv v(\Phi) \wedge h(\Psi) + v(\Psi) \wedge h(\Phi)  \, .
\end{equation}
On the other hand, in lemma \ref{lemma-null} we have proven that a
solution which satisfies $v(Z,Z)=0$ is, necessarily, a Kerr-NUT
metric. Note that the Killing vector $Z$ is orthogonal to the null
vector $v(Z)$, and using the results in \cite{fsEM-sym} it is easy
to prove that another Killing vector exists with the same property.
Consequently, these solutions have null orbits. Moreover, $v(Z,Z)=0$
is equivalent to $v(\Psi, \Psi) =0$. Thus we have:

\begin{proposition} \label{proposition-regular}
In the regular class {\rm D}$[1 1, 1 1]$ ($v(\Psi)\not= 0$,
$h(\Psi)\not=0$) we can distinguish three subclasses which can be
characterized by the following conditions:
\begin{enumerate}
\item[] \hspace*{-5mm} {\rm D}$[1 1, 1 1]_C$ ({\it Regular $C$-metrics}):
$N \not= 0$.
\item[] \hspace*{-5mm}  {\rm D}$[1 1, 1 1]_{K}$ ({\it Regular Kerr-NUT metrics with
non null orbits}): $N = 0$, $v(\Psi, \Psi) \not=0$.
\item[] \hspace*{-5mm}  {\rm D}$[1 1, 1 1]_{n}$ ({\it Solutions with null
orbits}): $v(\Psi, \Psi)=0$.
\end{enumerate}
\end{proposition}

We can easily relate prior classifications with ours. Classes I, II
III and IV by Edgar et al. \cite{edgar-alfonso} correspond to
specific classes or types of our approach: class I  to our type {\rm
D}$[\cdot 1, 0 0]$, class II  to our class {\rm D}$[1 1, 1
1]_{n}$,  class IIIA to our class {\rm D}$[0 1, 0 1]$, class IIIB
to our class {\rm D}$[1 1, 1 1]_C$, class IIIC to our class  {\rm
D}$[1 1, 1 1]_{K}$, and class IV  to our type  {\rm D}$[0 0, \cdot
1]$.

Finally, we present our results on the classification of the type
 D vacuum solutions in an algorithmic form by using a flow chart.
We use the following Weyl concomitants: the projectors on the
principal planes $v = \frac{1}{2} \, g + {\cal U} \cdot \bar{\cal
U}$ and $h = g-v$,  the linear first order vectors $\Phi$ and $\Psi$
given in (\ref{Sigma-Phi-Psi}) (or, equivalently, in
(\ref{rainich})), and the quadratic first order 2-form $N$ given in
(\ref{N-KN}). The explicite expression of $w$ and ${\cal U}$ in
terms of the Weyl tensor are given by \cite{fms}:
$$w = - \frac{{{\cal W}_{\alpha \beta}}^{\mu \nu} \
{{\cal W}_{\mu \nu}}^{\epsilon \delta} \, {{\cal W}_{\epsilon
\delta}}^{\alpha \beta}}{2 \  {{\cal W}_{\alpha \beta}}^{\mu \nu}
{{\cal W}_{\mu \nu}}^{\alpha \beta}} \, ; \qquad  {\cal U} =   \,
\frac{ {{\cal Q}_{\alpha \beta}}^{ \mu \nu} F_{\mu
\nu}}{\sqrt{3\,{{\cal Q}_{\alpha \beta}}^{ \mu \nu} F_{\mu \nu}
F^{\alpha \beta}}} \, ,  \quad  {\cal Q}\equiv \frac{1}{w} {\cal W}
- {\cal G} \, ,$$
where $F$ is an arbitrary 2-form such that ${{\cal Q}_{\alpha \beta}}^{ \mu
\nu} F_{\mu \nu} \not= 0$.



 \setlength{\unitlength}{0.8cm} {\footnotesize
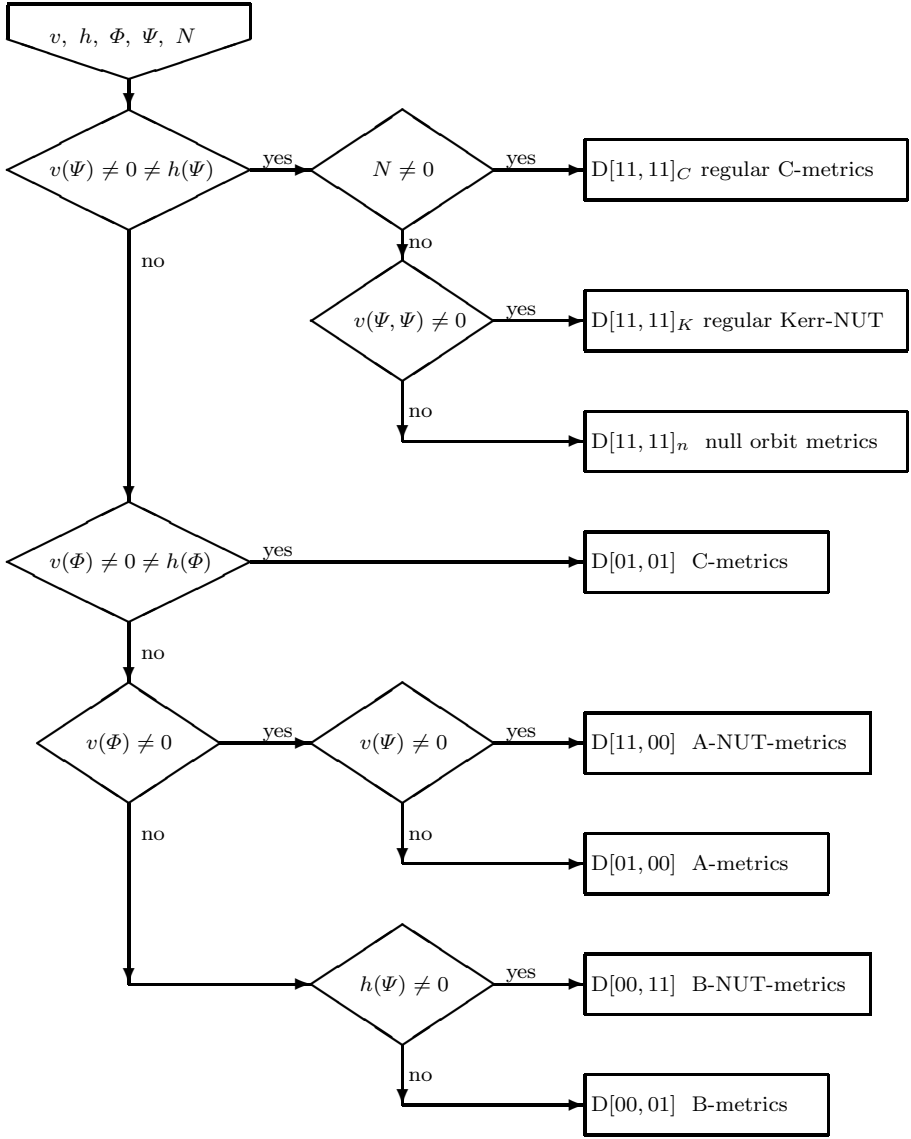
\noindent
\begin{picture}(12,20)(1,0)
\thicklines

\put(3,18){\line(3,1){2}}
 \put(3 ,18 ){\line(-3,1){2}}

\put(1.01,18.68){\line(0,1){0.55}}\put(5,19.25){\line(-1,0){4}}
 \put(5,19.25){\line(0,-1){0.57}} \put(3,18){\vector(0,-1){0.5}}

\put(1,16.5){\line(2,1){2}}\put(1,16.5){\line(2,-1){2}}
\put(5,16.5){\line(-2,1){2}} \put(5,16.5){\line(-2,-1){2}}

\put(1.7,18.6){$  v,\ h, \  \Phi, \ \Psi, \ N$}

\put(1.7,16.4){$  v ( \Psi) \neq 0 \neq h(\Psi)$}

\put(5,16.5){\vector(1,0){1}}

\put(6 ,16.5){\line(3,2){1.5}}\put(6 ,16.5){\line(3,-2){1.5}} \put(9
,16.5){\line(-3,2){1.5}} \put(9 ,16.5){\line(-3,-2){1.5}}

\put(7,16.4){$  N \neq 0$}

\put(9,16.5){\vector(1,0){1.5}}

\put(10.5,16){\line(1,0){5.3}} \put(10.5,16){\line(0,1){1}}
\put(15.8,17){\line(-1,0){5.3}} \put(15.8,17){\line(0,-1){1}}

\put(10.6,16.4){D$ [11,11]_C \    \textrm{regular  C-metrics}$}


\put(7.5,15.5){\vector(0,-1){0.5}}

\put(6 ,14){\line(3,2){1.5}}\put(6 ,14){\line(3,-2){1.5}} \put(9
,14){\line(-3,2){1.5}} \put(9 ,14){\line(-3,-2){1.5}}

\put(10.5,13.5){\line(1,0){5.3}} \put(10.5,13.5){\line(0,1){1}}
\put(15.8,14.5){\line(-1,0){5.3}} \put(15.8,14.5){\line(0,-1){1}}
\put(9,14){\vector(1,0){1.5}}

\put(7.5,13){\vector(0,-1){1}}

\put(10.5,11.5){\line(1,0){5.3}} \put(10.5,11.5){\line(0,1){1}}
\put(15.8,12.5){\line(-1,0){5.3}} \put(15.8,12.5){\line(0,-1){1}}
\put(7.5,12){\vector(1,0){3}}

\put(3,15.5){\vector(0,-1){4.5}}


\put(5,10){\vector(1,0){5.5 }}

\put(10.5,9.5){\line(1,0){4}} \put(10.5,9.5){\line(0,1){1}}
\put(14.5,10.5){\line(-1,0){4}} \put(14.5,10.5){\line(0,-1){1}}

\put(1.5,7  ){\line(3,2){1.5}}\put(1.5 ,7 ){\line(3,-2){1.5}}
\put(4.5 ,7){\line(-3,2){1.5}} \put(4.5 ,7 ){\line(-3,-2){1.5}}

\put(6,7  ){\line(3,2){1.5}}\put(6 ,7 ){\line(3,-2){1.5}} \put(9
,7){\line(-3,2){1.5}} \put(9 ,7 ){\line(-3,-2){1.5}}

\put(10.5,6.5){\line(1,0){4.7}} \put(10.5,6.5){\line(0,1){1}}
\put(15.2,7.5){\line(-1,0){4.7}} \put(15.2,7.5){\line(0,-1){1}}

\put(10.5,4.5){\line(1,0){4}} \put(10.5, 4.5){\line(0,1){1}}
\put(14.5,5.5){\line(-1,0){4}} \put(14.5,5.5){\line(0,-1){1}}

\put(6,3  ){\line(3,2){1.5}}\put(6 ,3 ){\line(3,-2){1.5}} \put(9
,3){\line(-3,2){1.5}} \put(9 ,3 ){\line(-3,-2){1.5}}

\put(10.5,2.5){\line(1,0){4.7}} \put(10.5,2.5){\line(0,1){1}}
\put(15.2,3.5){\line(-1,0){4.7}} \put(15.2,3.5){\line(0,-1){1}}

\put(10.5,0.5){\line(1,0){4}} \put(10.5, 0.5){\line(0,1){1}}
\put(14.5,1.5){\line(-1,0){4}} \put(14.5,1.5){\line(0,-1){1}}

\put(4.5,7){\vector(1,0){1.5 }} \put(9,7){\vector(1,0){1.5 }}

\put(3,3){\vector(1,0){3 }} \put(9,3){\vector(1,0){1.5 }}

\put(3, 6){\vector(0,-1){3}}

 \put(3, 9){\vector(0,-1){1}}

 \put(7.5, 6){\vector(0,-1){1}}
\put(7.5,5){\vector(1,0){3 }}

\put(7.5, 2){\vector(0,-1){1}}

\put(7.5,1){\vector(1,0){3 }}

\put(1.7,9.9){$  v ( \Phi) \neq 0 \neq h(\Phi)$} \put(6.7,13.9){$
v(\Psi, \Psi) \neq 0$}

\put(1,10){\line(2,1){2}}\put(1,10){\line(2,-1){2}}
\put(5,10){\line(-2,1){2}} \put(5,10){\line(-2,-1){2}}

\put(2.3,6.9){$ v(\Phi) \neq 0$} \put(6.8,6.9){$ v(\Psi) \neq 0$}

\put(6.8,2.9){$h(\Psi) \neq 0$}

\put(10.6,11.9){D$ [11,11]_n$\ \ null orbit metrics}

\put(10.6,13.9){D$ [11,11]_K$\  regular Kerr-NUT}
 \put(10.6,9.9){D$ [01,01]$\ \ C-metrics}

\put(10.6,6.9){D$ [11,00]$\ \  A-NUT-metrics}

\put(10.6,4.9){D$ [01,00]$\ \  A-metrics} \put(10.6,2.9){D$[00,11]$\
\  B-NUT-metrics} \put(10.6,0.9){D$[00,01]$\ \  B-metrics}

\put(7.6,1.4){no} \put(7.6,5.4){no} \put(7.6,15.2){no}
\put(7.6,12.4){no} \put(3.2,14.9){no}
 \put(3.2,8.4){no}   \put(3.2,5.4){no}
\put(5.2,16.6){yes}

\put(5.2,10.1){yes} \put(5.2,7.1){yes} \put(9.2,16.6){yes}
\put(9.2,14.1){yes} \put(9.2,7.1){yes} \put(9.2,3.1){yes}
\end{picture}

}


\section{Discussion and comments}
\label{discussion}

The Cartan invariant scheme based on the Riemann tensor and its
covariant derivatives was introduced by Brans \cite{brans} in
general relativity and, after Karlhede's work \cite{karlhede}, this
approach became more helpful within the relativistic framework.  The
Cartan-Karlhede method is based on working in an orthonormal (or a
null) frame, fixed by the underlying geometry of the Riemann tensor.
Nevertheless there are a lot of historic results which show that
the determination of a Riemann canonical frame is not always
necessary to label a family of metrics: theorems that characterize
locally flat spaces, Riemann spaces with a maximal group of
isometries, and locally conformally flat spaces are some examples.
Also the well known characterizations of the Stephani Universes or
of the Friedmann-Lema\^itre-Robertson-Walker Universes. These
examples show that the characterization conditions can involve
tensorial concomitants whereas the Cartan-Karlhede scheme only uses
scalar concomitants.

An example where the Cartan-Karlhede approach has shown its efficacy
is the study of the covariant derivatives of the Riemann tensor for
the type  D vacuum metrics presented by $\textrm{\AA}$man
\cite{aman} and  performed in \cite{edgar-alfonso}. But the results
we present here show that a tensorial approach can
bring new knowledge on this topic. Several identities with no clear
sense and which have previously been acquired with hard computer
support have been obtained here in an easy way and their plain
geometric meaning has been outlined.

In studying the Karlhede upper bound, we have used the invariant
Killing vector $Z$ and its associated Killing 2-form $\nabla Z$, and
only three different cases have had to be considered. For the
Kerr-NUT vacuum metrics, we obtain the expression
(\ref{nablaZ-Kerr-NUT}) which give $\nabla Z$ in terms of first
order invariants, and we arrive to theorem \ref{theorem-1}. For the
regular C-metrics, expressions (\ref{Killing-2-form-b}, \ref{m-Cg-1},
\ref{m-Cg-2}) give $\nabla Z$ in terms of first order invariants,
and we arrive to theorem \ref{theorem-2}. Finally, for the strict
Ehlers and Kundt C-metrics, expressions (\ref{Killing-2-form-b},
\ref{K}) show that $\nabla Z$ depend on first order invariants and
on a second order constant invariant $K$, and we arrive to theorem
\ref{theorem-3}.

These three theorems show that the Karlhede upper bound is two.
Nevertheless, there is a relevant difference between the third case
and the other ones. Theorems \ref{theorem-1} and \ref{theorem-2}
imply that a specific Kerr-NUT solution or a specific regular
C-metric can be characterized by exclusively using first order Weyl
concomitants. Our local intrinsic labeling of the Schwarszchild
\cite{fsS} and Kerr \cite{fsKerr} black holes are examples of this
fact. In particular, the {\it mass} and the {\it angular momentum}
are first order constant invariants. However, theorem
\ref{theorem-3} states that in order to distinguish two different
strict C-metrics, we need to calculate the second order constant
invariant $K$.

It is worth pointing out that the ${\cal D}$-metrics (charged
counterpart of the type  D vacuum solutions) also admit the
invariant complex Killing vector $Z$. Thus, an invariant approach to
these solutions similar to that presented here could provide a
better understanding of them. Note that in this case the (non
vanishing) Ricci tensor will also be an important piece in the
invariant analysis. In \cite{fsEM-sym}, \cite{fs-EM-align} we have
studied invariant properties of this type of metrics, which will be
useful for this forthcoming work.

This procedure could be also useful in studying the Cartan-Karlhede upper bound not only in type D solutions but also in solutions with another Petrov-Bel type. For example a similar invariant vector can be defined in type II metrics, and in type N and type III spacetimes where all the Weyl invariants vanish, the first derivatives of the Weyl tensor also define invariant vectors.

A generic classification of type  D space times is a powerful tool
in learning geometric properties of solutions to Einstein equations.
In \cite{fsD} we introduced a classification based on first order
Weyl constraints and where the principal planes play a symmetric
role: if a class is defined by a property of the time-like plane,
then there exists a space-like counterpart class. We find this fact
in the pioneer paper by Ehlers and Kundt \cite{ehku} where the
B-metrics are the time-like counterpart of the (space-like)
A-metrics. In \cite{fsD} we used our classification to label the A,
B and C charged metrics. Here we apply this approach to the full set
of type  D vacuum solutions and only six classes survive: the
Ehlers and Kundt A and B-metrics, their NUT generalization (the
A-NUT and B-NUT metrics), the Ehlers and Kundt C-metrics, and the
most regular class which includes three subclasses, the regular
Kerr-NUT metrics, the regular generalized C-metrics, and the
solutions with null orbits.

Our classification has a hierarchic structure with five possible
levels of degeneration when the Cotton tensor vanishes. In the
vacuum case only the first, the third and the fourth levels remain.
This hierarchic structure enables a simple operational
algorithm to be build to distinguish every class.

It is worth remarking that the generic character of our classification allows us to apply it to the full set of type D metrics or to any specific family of type D solutions. For example it has been implemented elsewhere \cite{fsD} to classify the charged counterpart of the (static) A, B and C-metrics and to achieve an algorithm to label every solution. In particular, an intrinsic characterization of the Reissner-Nordstr\"om has been obtained. A similar approach could be accomplished for the charged counterpart of the six classes of type D vacuum solutions considered here.

On the other hand, the generalization of this approach to other Petrov-Bel solutions is also possible. Indeed, types III and II admit again a privileged 2+2 almost-product structure defined by the two null Debever directions in type III and by a non-null Weyl eigenbivector in type II.

\section*{Appendix A:  Proof of lemmas \ref{lemma-ZZ-nonnull}, \ref{lemma-null}
and \ref{lemma-Cmetrics}} \label{A-lemmas}

\subsection*{A.1  Proof of lemma \ref{lemma-ZZ-nonnull}}

Suppose that $(Z,Z)=0$. If we differentiate this condition and take
into account (\ref{killing-2-form-c}) we obtain:
\begin{equation} \label{ZZ}
2 w^{\frac23} Z = [\bar{w}^{\frac13} (Z, \bar{Z}) - 2 m] \Pi(Z) \, .
\end{equation}
From here we have $Z \wedge \Pi(Z) = 0$, that is, $Z$ is a null
vector which lies on the time-like principal plane: $Z = \Pi(Z)$.
Then (\ref{x-identities-a}) implies $Z \wedge \bar{Z} =0$ and
consequently $(Z, \bar{Z})=0$, and \ref{ZZ} is equivalent to:
\begin{equation} \label{ZZ-b}
Z = \Pi(Z) \, , \qquad m + w^{\frac23} = 0 \, .
\end{equation}
Now if we differentiate the first constraint in (\ref{ZZ-b}) (or
equivalently $i(Z){\cal U} = i(Z)\bar{\cal U}$), and take into
account (\ref{killing-2-form-c}) and $Z \wedge \bar{Z} =0$, we
arrive at $m  = w^{\frac23}$ which is not compatible with the second
constraint in (\ref{ZZ-b}).

\subsection*{A.2  Proof of lemma \ref{lemma-null}}
\label{A-null}

Condition $v(Z,Z)=0$ equivalently states $(Z,Z) + \Pi(Z,Z) =0$. If
we differentiate this scalar condition and we take into account the
expression of the covariant derivatives (\ref{killing-2-form-c}) of
$Z$ and (\ref{dw-dU}) of ${\cal U}$, we obtain the following
expression for $m$:
\begin{equation} \label{m-null}
m = - \frac12 \bar{w}^{\frac13} (Z, \bar{Z}) - \nu \,  , \quad  \nu
\equiv w^{\frac23} + \frac12 w^{\frac13} (Z,Z) \, .
\end{equation}
Then equation (\ref{m-eq}) becomes:
\begin{equation} \label{nu-h}
\nu h(\bar{Z}) + \bar{\nu} h(Z) = 0 \, .
\end{equation}
On the other hand, if we differentiate (\ref{m-null}) and make use
of (\ref{dw-dU}) and (\ref{killing-2-form-c}) we obtain a new scalar
condition:
\begin{equation} \label{m-null-b}
(Z, \bar{Z}) + 2m \bar{w}^{-\frac13} + 2 \bar{m} w^{-\frac13} = 0 \,
.
\end{equation}
Finally, if we differentiate this equation and we take into account
(\ref{dw-dU}) and (\ref{killing-2-form-c}) we arrive at:
\begin{equation} \label{nu-v}
\nu v(\bar{Z}) + \bar{\nu} v(Z) = 0 \, .
\end{equation}
Constraints (\ref{nu-h}) and (\ref{nu-v}) imply $\nu \bar{Z} +
\bar{\nu} Z = 0$. Consequently $Z \wedge \bar{Z}= 0$ or $\nu=0$.
This last condition and (\ref{m-null}) and (\ref{m-null-b}) lead to
$(Z, \bar{Z})=0$. This condition, hypothesis $(Z,Z) + \Pi(Z,Z) =0$,
and identity (\ref{x-identities-a}) imply $\Pi(Z) = Z$, which also
implies, with (\ref{x-identities-a}), $Z \wedge \bar{Z} = 0$.

Note that (\ref{m-null}) gives the scalar $m$ in terms of 0th- and
1st-order Riemann derivatives for the type D vacuum solutions
satisfying $v(Z,Z)=0$. Thus, we could state for them a specific
theorem similar to theorems \ref{theorem-1} and \ref{theorem-2}.
Nevertheless we prefer use this lemma \ref{lemma-null} and consider
this case as included in theorem \ref{theorem-1}.

\subsection*{A.3  Proof of lemma \ref{lemma-Cmetrics}}

From the hypothesis $Z \wedge \bar{Z}\not=0$ we have necessarily $Z
\wedge \Pi(Z)\not=0$. Then, condition $Z \wedge \Pi(\bar{Z})=0$
implies that (\ref{m-eq}) is equivalent to:
\begin{equation} \label{Cmetrics-a}
\bar{\mu} Z = \mu \Pi(\bar{Z}) \, , \qquad  \bar{\nu} Z = \nu \Pi(\bar{Z}) \, .
\end{equation}
On the other hand, if we differentiate $Z \wedge \Pi(\bar{Z})=0$ and
we make use of (\ref{dw-dU}) and (\ref{killing-2-form-c}) we obtain
a tensorial equation. Its trace leads to:
\begin{equation} \label{Cmetrics-b}
(\bar{m} + \bar{\mu}) Z = (m + \mu) \Pi(\bar{Z}) \, , \quad
(\bar{w}^{\frac23} + 2 \bar{\nu}) Z = ( w^{\frac23} + 2 \nu)
\Pi(\bar{Z}) \, ,
\end{equation}
where in obtaining the second equation we have used the fact that $Z \wedge
\Pi(\bar{Z})=0$ implies $\Pi(Z, \bar{Z}) Z = (Z,Z) \Pi(\bar{Z})$ and
$\Pi(Z, \bar{Z})  \Pi(\bar{Z}) = (\bar{Z},\bar{Z}) Z$. If we again
make use of these relations, from equations (\ref{Cmetrics-a})
(\ref{Cmetrics-b}) we obtain:
\begin{eqnarray} \label{Cmetrics-c}
\bar{m} Z = m \Pi(\bar{Z}) \, , \qquad \bar{w}^{\frac23} Z =
w^{\frac23} \Pi(\bar{Z}) \, , \label{Cmetrics-c} \\[2mm]
(Z, \bar{Z})V= 0 \, , \quad  \Pi(Z, \bar{Z}) V = 0 \, , \quad V
\equiv w^{\frac13} Z - \bar{w}^{\frac13} \Pi(\bar{Z}) \, .
\label{Cmetrics-d}
\end{eqnarray}
Under the hypothesis $Z \wedge \bar{Z}\not=0$, at least one of the
scalars $(Z, \bar{Z})$ and $\Pi(Z, \bar{Z})$ does not vanish.
Consequently (\ref{Cmetrics-d}) implies $V=0$. This constraint and
(\ref{Cmetrics-c}) lead to $\bar{m} = m$, $\bar{w} =w$ and
$\Pi(\bar{Z}) = Z$. Moreover, the solution is a strict C-metric
because it has real Weyl eigenvalues.

\section*{Appendix B: Proof of lemma \ref{lemma-Phi-Psi}}
\label{B-lemmas}

The conditions involved in lemma \ref{lemma-Phi-Psi} can be stated
by using the projections $v(\chi)$ and $h(\chi)$ of complex vector
$\chi = \frac12 [\Phi + \ci \Psi]$. In terms of $\{w, {\cal U}, Z\}$
these projections take the expression:
\begin{equation} \label{v-chi-Z}
 2 v(\chi) = w^{\frac13}[i(Z)({\cal U}) +
i(Z)(\bar{{\cal U}})] \, , \qquad  2 h(\chi) =
w^{\frac13}[i(Z)({\cal U}) - i(Z)(\bar{{\cal U}})] \, .
\end{equation}

Suppose that $v(\Phi) = 0$, that is, $v(\chi) = - v(\bar{\chi})$. If
we calculate the covariant derivative of this equation and we take
into account expressions (\ref{v-chi-Z}) and derivatives
(\ref{dw-dU}) and (\ref{killing-2-form-c}) we obtain a 2-tensorial
equation $E_{\alpha \beta}= 0$. The total projection of this
equation on the time-like plane, $v^{\lambda \alpha} v^{\mu \beta}
E_{\alpha \beta}= 0$, leads to $v(\Psi) \otimes v(\Psi) = 0$, and so
$v(\Psi) = 0$. Consequently point (i) is proven.

Suppose now that $v(\Psi) = 0$, that is, $v(\chi) = v(\bar{\chi})$.
If we calculate the covariant derivative of this equation and we
take into account expressions (\ref{v-chi-Z}) and derivatives
(\ref{dw-dU}) and (\ref{killing-2-form-c}) we obtain a 2-tensorial
equation $F_{\alpha \beta}= 0$. The mixed projection of this
equation on the time-like and space-like planes, $v^{\lambda \alpha}
h^{\mu \beta} F_{\alpha \beta}= 0$, leads to $v(\Phi) \otimes
h(\Psi) = 0$, and so either $v(\Phi) = 0$ or $h(\Psi) = 0$.
Consequently point (iii) is proven.

The proof of points (ii) and (iv) of lemma \ref{lemma-Phi-Psi} is
similar to the proof of points (i) and (iii) by exchanging $v$ for
$h$.

\begin{acknowledgements}
 This work has been supported by the Spanish ``Ministerio de
Econom\'{\i}a y Competitividad", MICINN-FEDER project FIS2012-33582.
\end{acknowledgements}



\end{document}